\documentclass[runningheads]{llncs}
\usepackage{tcolorbox}
\usepackage{tabularx}
\usepackage[framemethod=tikz]{mdframed}
\usepackage{multirow}
\usepackage{subcaption}
\usepackage{booktabs} % For formal tables
\usepackage{array}

\newcommand{\airtown}{\texttt{AirTOWN}}

\mdfdefinestyle{AssumptionBox}{
    linecolor=blue,
    outerlinewidth=2pt,
    roundcorner=10pt,
    innertopmargin=10pt,
    innerbottommargin=10pt,
    innerrightmargin=10pt,
    innerleftmargin=10pt,
    backgroundcolor=gray!20
}

\usepackage{booktabs}
\usepackage{makecell}
\usepackage{tabularx}
\usepackage{booktabs}
\usepackage{booktabs}
\usepackage{amsmath}

\usepackage{hyperref}
\usepackage{tikz}

\usetikzlibrary{shapes.geometric, arrows, positioning}

\usetikzlibrary{shapes.geometric, arrows.meta, chains, positioning, quotes}

\definecolor{lightgray}{gray}{0.9}

\usepackage[T1]{fontenc}
\usepackage{graphicx}
\begin{document}
\title{AirTOWN: A Privacy-Preserving Mobile App for
Real-time Pollution-Aware POI Suggestion}
\titlerunning{AirTOWN: A Privacy-Preserving Mobile App}
\author{Giuseppe Fasano\orcidID{0009-0000-7249-9726} \and Yashar Deldjoo\orcidID{0000-0002-6767-358X} \and Tommaso Di Noia\orcidID{0000-0002-0939-5462}}
\authorrunning{G. Fasano et al.}
% First names are abbreviated in the running head if there are more than two authors, 'et al.' is used.

\institute{Politecnico di Bari, Bari, Italy \\
\email{firstname.lastname@poliba.it}}
%giuseppe.fasano@poliba.it, yashar.deldjoo@poliba.it, tommaso.dinoia@poliba.it}}

\maketitle              % typeset the header of the contribution

\vspace{-6mm}
\begin{abstract}
This demo paper presents \airtown, a privacy-preserving mobile application that provides real-time, pollution-aware recommendations for points of interest (POIs) in urban environments. By combining real-time Air Quality Index (AQI) data with user preferences, the proposed system aims to help users make health-conscious decisions about the locations they visit. The application utilizes collaborative filtering for personalized suggestions, and federated learning for privacy protection, and integrates AQI data from sensor networks in cities such as Bari, Italy, and Cork, UK. In areas with sparse sensor coverage, interpolation techniques approximate AQI values, ensuring broad applicability. This system offers a poromsing, health-oriented POI recommendation solution that adapts dynamically to current urban air quality conditions while safeguarding user privacy.

%\keywords{First keyword  \and Second keyword \and Another keyword.}
\end{abstract}
\vspace{-11mm}

\section{Introduction}
Air pollution is a global challenge, with 99\% of the population breathing air exceeding WHO guideline limits \cite{who_airpollution}. This issue is particularly acute in urban areas due to vehicle emissions and industrial processes. Point-of-interest (POI) recommendation systems offer a unique opportunity to enhance user experiences and support public health by incorporating health-conscious suggestions.

This paper introduces \textit{\airtown}, a mobile application that integrates real-time pollution data with user preferences to provide tailored POI recommendations for health-conscious urban navigation. The system emphasizes \textit{personalization}, \textit{pollution awareness}, \textit{real-time sensory data integration}, \textit{privacy preservation}, and considers the \textbf{distance} to venues. \airtown~helps users make healthier choices by identifying nearby POIs, personalizing suggestions using collaborative filtering (CF), and re-ranking them based on Air Quality Index (AQI) data from AirSence\footnote{\url{https://airsence.com/}}. 

To ensure accuracy, \airtown~uses sensor networks in Bari, Italy, and Cork, UK, providing AQI, temperature, humidity, and pressure data. These cities represent diverse urban settings with varying air quality challenges. In areas with limited sensor coverage, interpolation techniques approximate AQI, extending the system’s applicability.

Existing systems often monitor exposure \cite{LIN2023114186} or predict pollution along routes without suggesting healthier POIs \cite{IoT-app}. Health-based recommendation systems focus on static health metrics \cite{health-aware,10.1145/3503252.3531312}, and privacy-centric solutions \cite{POIRSDemo,chen2020practical} lack real-time environmental integration. In contrast, \airtown~combines real-time pollutant data, personalization, and privacy to offer dynamic, health-conscious recommendations. While there are limitations, \airtown~represents a significant step toward supporting healthier urban navigation through real-time data integration and privacy-conscious design.

\begin{figure}[!t]
    \centering
    \includegraphics[width=0.850\linewidth]{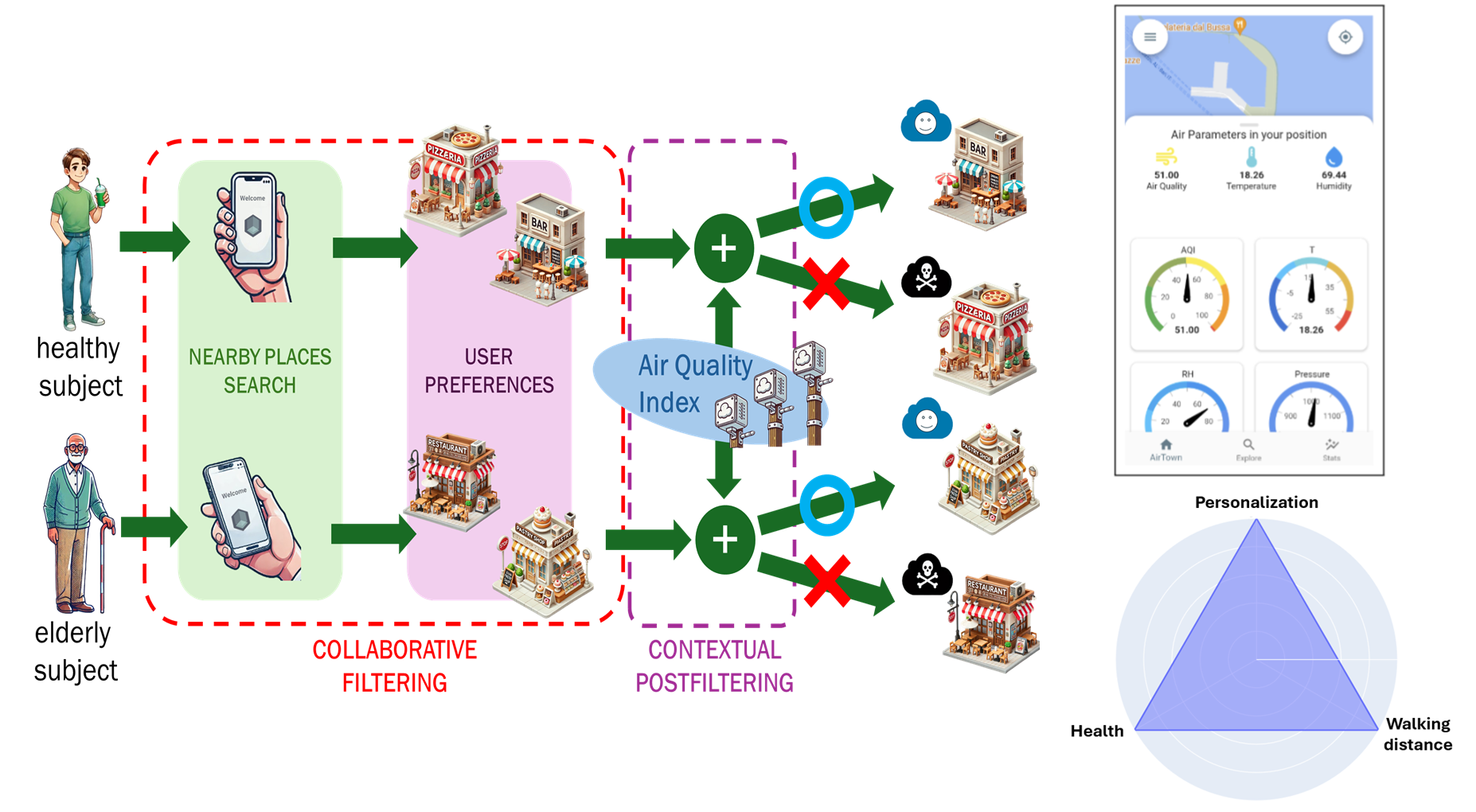}
    \caption{Recommendation process in \airtown}
    \label{POI_recsys_scheme}
\end{figure}

\section{System Architecture}
% The architecture of the \airtown app is designed to leverage a multi-layered client-server model to enhance user experience by providing personalized point-of-interest (POI) recommendations based on air quality and user preferences. This section describes the layers involved in the system's architecture, as represented in Figure \ref{FL_architecture}.
%As Figure \ref{FL_architecture} shows, 
The architecture of \airtown~is designed to leverage a client-server model featuring four layers: Application, Service, Interface, and Data Resources Layer.

%\begin{figure} [h]
    %\centering
    %\includegraphics[width=1\linewidth]{images/FL architecture scheme condensed.png}
    %\caption{The architecture of the Federated Recommender System in \airtown}
    %\label{FL_architecture}
%\end{figure}

\textbf{Application Layer.} This layer serves as the user interface on mobile smartphones and performs POI suggestions integrating air quality information. To obtain an effective POI recommendation system (RS), we focused on balancing three key factors: personalization, pollution levels, and distance of the target venue (POI). When a user requests suggestions, the system first identifies POIs within a certain radius (e.g., 1 km) of the user’s current location, addressing the distance factor. Next, a collaborative filtering (CF) model predicts user preferences to personalize the recommendations. We chose the Matrix Factorization (MF) solution over more sophisticated alternatives as Deep Learning, because of its better transparency and lower complexity, allowing for distributed training on devices with limited resources. Finally, the system applies a weighted re-ranking function that balances the predicted preferences and the AQI of the POIs. The parameter $\alpha \in [0,1]$ modulates the strength of AQI influence in the re-ranking process $S = \alpha \cdot S_{MF} + (1 - \alpha) \cdot S_{AQI}$. To enhance MF performances, the model is trained on user preferences, which are locally collected via surveys on already visited POIs. The Application Layer never shares user data and the update of the local model is performed in a client-server implementation of FL paradigm, allowing for user privacy.

\textbf{Service Layer.} Hosted on a server, the Service Layer manages back-end processes such as login, registration, and data retrieval for the Application Layer. It approximates AQI data for unsensed locations using radial basis function interpolation and performs federated learning rounds for global model training via the Federated Averaging protocol. Only item embedding updates are shared, while user embeddings remain local.

\textbf{Interface and Data Resources Layer.} This layer comprises APIs linking the Service Layer with data resources. The AirSENCE API provides localized air quality data, while Directions and Places APIs enable navigation and location information. The Google Database manages Google-related data (Photos, place information, routing data), and the AirSENCE Database stores air quality measurements and federated model parameters.

\vspace{-4mm}
%\section{Experiments and Demonstrations.}
%To show the functioning of \airtown, we provide intra-user and inter-user investigations in Bari (Italy). To overcome the current limited sensor coverage provided by AirSENCE over the city, we considered synthetic AQI data.\footnote{\url{https://anonymous.4open.science/r/Airtown-Application/}}

\section{Intra-user and Inter-user Demonstration}

To evaluate the effectiveness of \texttt{AirTOWN} in providing personalized, health-conscious recommendations, we conducted intra-user and inter-user demonstrations in Bari, Italy, using synthetic AQI data due to limited sensor coverage. This setup enabled us to simulate varying air quality scenarios and assess the impact of recommendation parameters.\footnote{\url{https://anonymous.4open.science/r/Airtown-Application/}}

Two users were simulated in Aldo Moro Square, Bari, each using \texttt{AirTOWN} on mobile devices. User preferences were collected to compute individual embeddings through local models, allowing for personalized recommendations. A virtual sensor grid structured as a $1 \times 1$ km layout surrounding each user provided AQI values randomly assigned within a range of 20 to 70, enabling realistic air quality variation. Both users requested restaurant recommendations, with User 1 representing a generally healthy individual and User 2 representing an elderly individual with increased sensitivity to air quality.

\subsection{Demonstration Results}

The demonstration involved adjusting the weighting parameter $\alpha$ to control the influence of AQI versus user preferences in the recommendation ranking.

\begin{figure}[h]
    \centering
    \includegraphics[width=0.750\linewidth]{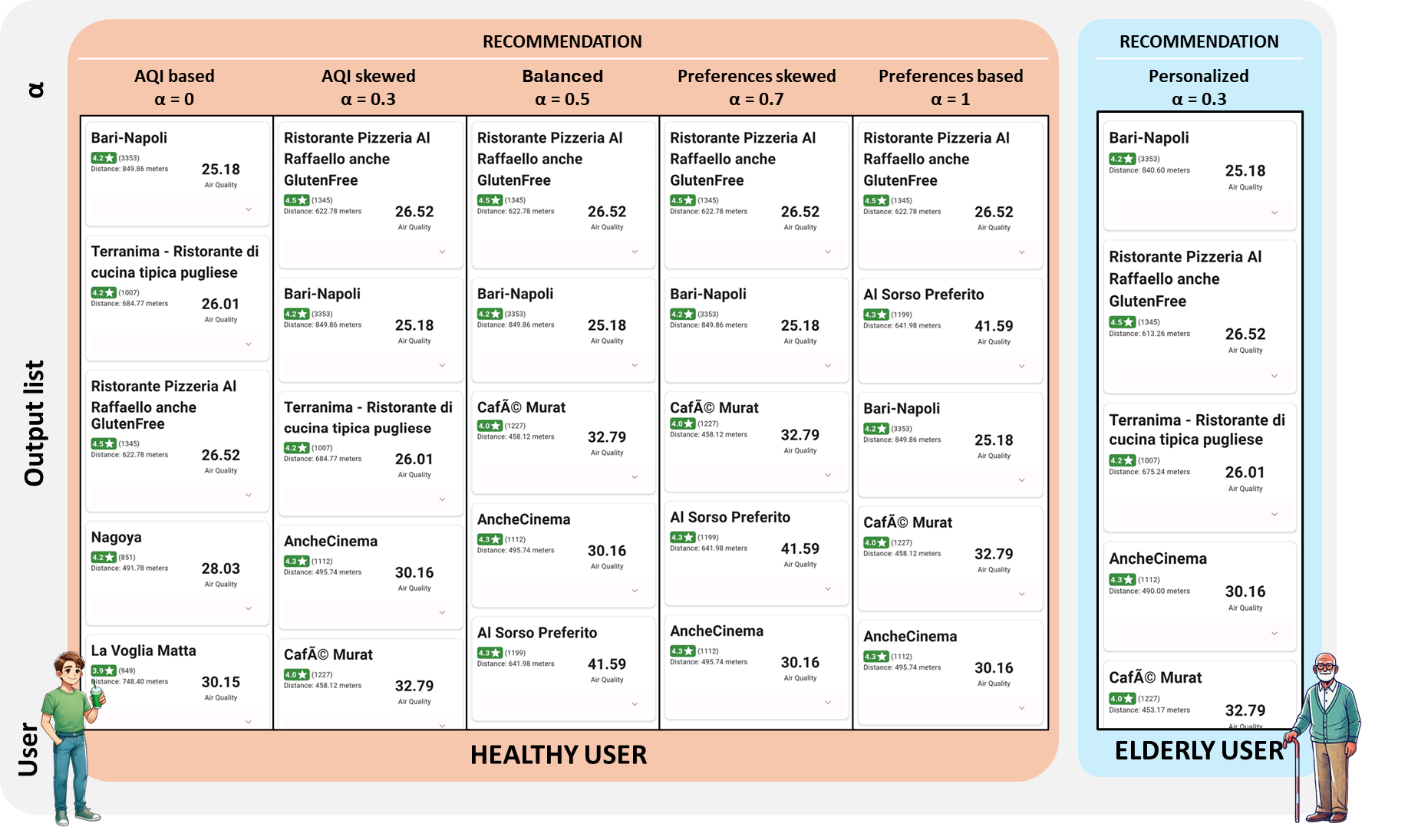}
    \caption{Recommendation lists for two users in the simulated scenario.}
    \label{simulations}
\end{figure}

\textbf{Intra-user demonstration.} As illustrated by the red-boxed results in Figure~\ref{simulations}, varying $\alpha$ shows different recommendation lists:
\begin{itemize}
    \item \textbf{$\alpha = 0$ (AQI-driven):} Recommendations are ordered solely by AQI, disregarding preferences.
    \item \textbf{$\alpha = 1$ (preference-driven):} Recommendations prioritize user preferences, sometimes listing POIs with higher AQI.
    \item \textbf{$\alpha = 0.5$ (balanced):} Recommendations are adjusted to reflect both AQI levels and personal tastes, achieving a compromise between health considerations and preferences.
\end{itemize}

\textbf{Inter-user demonstration.} User 2, representing an elderly individual, applied $\alpha = 0.3$ to favor lower AQI levels due to increased health sensitivity, as shown in the blue-boxed list in Figure~\ref{simulations}. This setting provided a recommendation list distinct from that of User 1, who prioritized preferences over AQI, demonstrating \texttt{AirTOWN}'s adaptability to individual health needs.

%\subsection{Conclusion}
%These demonstrations validate \texttt{AirTOWN}'s capability to balance personal preferences and air quality data for health-conscious, tailored POI recommendations. Although limited sensor coverage necessitated synthetic data, results indicate the system's potential in urban navigation with a focus on user well-being. Future enhancements will focus on expanding real-time sensor networks and refining AQI interpolation for improved accuracy in diverse urban settings.

%\textbf{Model and Setting Preparation.} For the demonstration, two users were considered. At the moment of the experimentation, each user was located in Aldo Moro Square, Bari, and had \airtown~installed on his own mobile. After collecting their preferences, each local model computed its user embedding and the recommendation system was ready. The user was located at the center of a virtual distribution of eight sensors, in a $1 \times 1$ km grid, and each sensor was initialized with a random AQI value $\in [20,70]$. Each user asked for a restaurant recommendation. We assume that the first user is a normal healthy user, while the second one is an elderly user.

%\textbf{Intra-user demonstration.} As the red box in Figure \ref{simulations} shows, the order of the first five suggested POIs changes according to $\alpha$ among the screenshots (The interface was slightly modified to enhance readability). When $\alpha = 0$, the recommendation is completely AQI-driven; indeed, the suggestions are sorted in descending AQI order. When $\alpha = 1$, the recommendation only rely on user preferences, and POIs with higher AQI can appear in higher positions, like "Al Sorso Preferito" (AQI = 41.59). When $\alpha = 0.5$, the same importance is given to AQI and user preferences, leading to meeting point of the previous ones.

%\textbf{Inter-user demonstration.} An elderly user should pay more attention to the air quality of the POIs, choosing $\alpha = 0.3$ for the suggestions. The first five suggestions for the second user are displayed in the blu box in Figure \ref{simulations}. His sequence of POIs differs from the respective one of the healthy user, because of the preferences differences among them.

\vspace{-7mm}

\section{Conclusion}
We presented \airtown, a mobile application that combines personalization, real-time pollution awareness, privacy preservation, and proximity considerations to deliver health-conscious point-of-interest (POI) recommendations. % By integrating collaborative filtering with federated learning, \airtown~provides users with tailored suggestions while protecting the privacy of personal data.
Initial experiments highlight the effectiveness of the Mobile App. in balancing user preferences with real-time air quality data, making it a valuable tool for urban navigation that promotes healthier choices. 
Future work will focus on the privacy aspect, involving solutions to protect the updates, such as differential privacy. Further evaluation of the App. effectiveness will be conducted with a larger user base.

Overall, this paper situates \textit{\airtown} within the broader context of trustworthy recommender systems \cite{deldjoo2022survey,deldjoo2024fairness,deldjoo2024understanding,deldjoo2025cfairllm,nazary2025poison} by emphasizing privacy preservation, transparency, and health-conscious decision-making. We have also plans to draw inspiration from recent advancements in generative recommender models \cite{biancofiore2024interactive,deldjoo2024review,deldjoo2024recommendation}, which demonstrate the potential of dynamic, user-adaptive recommendations.

%While current limitations, such as limited sensor coverage and reliance on simpler interpolation techniques, offer areas for improvement, future work will focus on expanding sensor networks and enhancing recommendation accuracy. Ultimately, \airtown~demonstrates a promising step toward urban health-focused applications that support user well-being and environmental awareness.
%\airtown is a user-friendly mobile application which combines personalization, pollution levels, distance, and privacy to provide effective and health-conscious POI recommendations. \airtown allows the user to explore new nearby places both involving personal preferences and real-time sensors data. 

% LIMITATIONS:
%- Simulated data
%- Airsence small coverage
%- A better model than RBF interpolation can be adopted 
%- A better re-rank function can be considered

% PROS: 
%- MF more thrustworthy than more complex methods?
%- MF lighter than ANN, better for smartphone computational power
%- MF more scalable

\begin{credits}
\subsubsection{\ackname} This work was partially supported by the following projects: PASSEPARTOUT, LUTECH DIGITALE 4.0, P+ARTS, 2022LKJWHC - TRex-SE: Trustworthy Recommenders for Software Engineers, 2022ZLL7MW - Conversational Agents: Mastering, Evaluating, Optimizing.

%\subsubsection{\discintname}
%It is now necessary to declare any competing interests or to specifically state that the authors have no competing interests. Please place the statement with a bold run-in heading in small font size beneath the (optional) acknowledgments\footnote{If EquinOCS, our proceedings submission system, is used, then the disclaimer can be provided directly in the system.}, for example: The authors have no competing interests to declare that are relevant to the content of this article. Or: Author A has received research grants from Company W. Author B has received a speaker honorarium from Company X and owns stock in Company Y. Author C is a member of committee Z.
\end{credits}

\bibliographystyle{splncs04}
\bibliography{refs}

\end{document}